\def\degrees{^\circ}

\def\arcsec{^{\prime\prime}}
\def\etal{{\it et al.}}
\def\kms{$\mathrm {km}~\mathrm s^{-1}$}
\def\om{\Omega_{\rm p}}
\def\len{a_B}
\def\lag{D_L}
\def\vpd{{\cal R}}
\def\kmsa{$\mathrm {km}\mathrm/s/\arcsec$}

\documentstyle[11pt,newpasp,twoside]{article}

\begin{document}

\title{The Bar Pattern Speed in NGC 7079}
\author{Victor P. Debattista}
\affil{Astronomisches Institut der Universit\"at Basel, Venusstrasse 7, 
CH-4102 Binningen, Switzerland}

\author{T.B. Williams}
\affil{Dept. of Physics and Astronomy, Rutgers University, PO Box 849, 
Piscataway, NJ  08855, USA}

\begin{abstract}
We have used 2D Fabry-Perot absorption-line spectroscopy of the SB0 galaxy
NGC 7079 to measure its bar pattern speed, $\om$.  As in all previous cases
of bar pattern speed measurements, we find a fast bar.  We estimate that 
NGC 7079 has been undisturbed for {\it at least} the past Gyr or roughly 8 bar
rotations, long enough for the bar to have slowed down significantly through 
dynamical friction if the disk is sub-maximal.
\end{abstract}

\section{Introduction}
The fundamental parameter in barred galaxy (SB) dynamics is their bar pattern 
speed, $\om$, which can be parametrized by the ratio $\vpd \equiv \lag/\len$,
where $\lag$ is the corotation radius and $\len$ is the bar semi-major axis.
A bar is termed fast when $1.0 \leq \vpd \leq 1.4$.  $N$-body simulations by 
Debattista \& Sellwood (1998), following the theoretical work of Weinberg 
(1985), showed that dynamical friction on a bar from an isotropic dark matter
halo preserves a fast bar only if the disk is very close to maximal, in the 
sense that the luminous disk provides most of the rotational support in the 
inner galaxy.  Tremaine \& Ostriker (1999) suggested that a flattened, 
rapidly rotating inner halo can reduce dynamical friction sufficiently to 
preserve fast bars even in sub-maximal disks, but Debattista \& Sellwood 
(2000) showed that the halo rotation required is large; the stellar halo 
of the Milky Way shows no evidence for such large rotation.

Indirect evidence for fast bars in SB galaxies comes from 
hydrodynamical models of gas flow, particularly at the shocks.  Three 
such studies are: $\vpd = 1.3$ in NGC 1365 (Lindblad \etal\ 1996), 
$\vpd = 1.3$ in NGC 1300 (Lindblad \& Kristen 1996), and $\vpd = 1.2$ for NGC 
4123 (Weiner \etal\ 2000, who also showed that a $80\%-100\%$ 
maximal disk is highly favored in NGC 4123).  A direct, model-independent 
method for measuring $\om$, developed by Tremaine \& Weinberg (1984) gives
$\om = \frac{1}{\sin i} \frac{<V_{los}-V_{sys}>}{<X>}$, where $(X,Y)$ are
galaxy-centered coordinates parallel and perpendicular to the disk's major
axis, and all averages are at fixed $Y$ and weighted by the luminosity.  
Using the Tremaine-Weinberg (TW) method with slit spectra, Merrifield \& 
Kuijken (1995) and Gerssen \etal \ (1998) found fast bars in NGC 936 
$(\vpd = 1.4 \pm 0.3)$ and NGC 4596 $(\vpd = 1.15^{+0.38}_{-0.23})$ 
respectively.  No slow bars ($\vpd > 1.4$) in high surface brightness 
(HSB) galaxies have been discovered to date.

\section{Observations and Results}
We used the TW method to measure $\om$ in NGC 7079, using a 
full-2D velocity field.  We observed NGC 7079 with the Rutgers 
Fabry-Perot imaging interferometer on the CTIO 4m telescope.  We used 
the CaII 8542.14 \AA \ absorption line, redshifted to 8618 \AA, scanning 
the spectrum from $8608$ \AA \ to $8631$ \AA, in steps of $1$ \AA, for 
a total of 25 exposures of 900 seconds each.  Voigt profiles were fitted 
to the spectrum at each pixel of the reduced images, giving maps of the 
velocity and dispersion.  Data extracted along the slit PAs of Bettoni 
\& Galletta (1997, BG97) match their data with a worst reduced 
$\chi^2 = 1.2$ for 15 degrees of freedom.  The rotation curve, which we 
then corrected for asymmetric drift, was extracted using a tilted ring fit.

We used ellipse fits to $U$, $B$, $V$, $R$ and $I$ 
exposures obtained at the CTIO 0.9m telescope to obtain 
$i = {49.87\degrees}^{+0.23}_{-0.25}$, PA $= 78.8\degrees\pm 0.1\degrees$ and 
$\len = 24.9\arcsec$.  The $B-I$ map shows near-constant color for the 
disk and bulge separately, indicating uniform (or zero) internal extinction.

Applying the TW equation, we obtain $\om \simeq 9.33 \pm 0.25$ \kmsa, which 
gives $\lag = 23 \arcsec \pm 3\arcsec$, so that $\vpd \simeq 0.9 \pm 0.1$, 
making this a fast bar.  NGC 7079 is a currently undisturbed, HSB galaxy, 
offset only 0.3 mag from the Tully-Fisher relation of Courteau \& Rix 
(1992).  The nearest galaxy, ESO 287-37, is at a projected separation
of 363 kpc; assuming a relative velocity of 300 \kms, their last closest 
approach would have been over 1 Gyr, or about 8 bar rotations, ago, long 
enough for a bar in a sub-maximal disk to have slowed down significantly.  
BG97 reported faint O[III] emitting gas out to $\sim 15\arcsec$ 
counter-rotating relative to the stars.  If it originated in an accretion 
event, this gas, being retrograde, could not have added angular momentum 
to the bar. NGC 7079 must therefore be a maximum disk.


\vspace*{-0.25cm}
\begin{references}
\reference Bettoni, D. \& Galletta, G. 1997, \aap, 124, 61
\reference Courteau, S., \& Rix, H.-W.\ 1999, \apj, 513, 561
\reference Debattista, V.P. \& Sellwood, J.A. 1998, \apj, 493, L5
\reference Debattista, V.P. \& Sellwood, J.A. 2000, \apj, 544, to appear
\reference Gerssen, J., Kuijken, K. \& Merrifield, M.R. 1999, \mnras, 306, 926
\reference Lindblad, P. A. B., \& Kristen, H. 1996, \aap, 313, 733
\reference Lindblad, P. A. B., Lindblad, P. O., \& Athanassoula, E. 1996, 
\aap, 313, 65
\reference Merrifield, M.R. \& Kuijken, K. 1995, \mnras, 274, 933
\reference Tremaine, S. \& Ostriker, J. P. 1999, \mnras, 306, 662
\reference Tremaine, S. \& Weinberg, M.D. 1984, \mnras, 282, L5
\reference Weinberg, M. D. 1985, \mnras, 213, 451
\reference Weiner, B.J., Sellwood, J.A. \& Williams, T.B. 2000, \apj, to appear
\end{references}
\end{document}